\documentclass[sigconf]{acmart}
\usepackage{multicol}
\usepackage{multirow}
\usepackage{pifont}
\usepackage{bbding}
\usepackage{subfigure}
\usepackage{balance}
\AtBeginDocument{%
  \providecommand\BibTeX{{%
    \normalfont B\kern-0.5em{\scshape i\kern-0.25em b}\kern-0.8em\TeX}}}

\copyrightyear{2023}
\acmYear{2023}
\setcopyright{acmlicensed}
\acmConference[SIGIR '23]{Proceedings of the 46th International ACM SIGIR Conference on Research and Development in Information Retrieval}{July 23--27, 2023}{Taipei, Taiwan}
\acmBooktitle{Proceedings of the 46th International ACM SIGIR Conference on Research and Development in Information Retrieval (SIGIR '23), July 23--27, 2023, Taipei, Taiwan}
\acmPrice{15.00}
\acmDOI{10.1145/3539618.3591853}
\acmISBN{978-1-4503-9408-6/23/07}

\begin{document}

\title{Exploring the Spatiotemporal Features of Online Food Recommendation Service}

% why we don't need self-attention in online food ordering service.


\author{Shaochuan Lin}
\affiliation{%
  \institution{Alibaba Group}
  \city{Hangzhou}
  \country{China}
}
\email{lin.lsc@alibaba-inc.com}

\author{Jiayan Pei}
\affiliation{%
  \institution{Alibaba Group}
  \city{Hangzhou}
  \country{China}
}
\email{jiayanpei.pjy@alibaba-inc.com}

\author{Taotao Zhou}
\affiliation{%
  \institution{Alibaba Group}
  \city{Shanghai}
  \country{China}
}
\email{taotao.zhou@alibaba-inc.com}

\author{Hengxu He}
\affiliation{%
  \institution{Alibaba Group}
  \city{Shanghai}
  \country{China}
}
\email{hengxu.hhx@alibaba-inc.com}

\author{Jia	Jia}
\affiliation{%
  \institution{Alibaba Group}
  \city{Shanghai}
  \country{China}
}
\email{jj229618@alibaba-inc.com}

\author{Ning Hu}
\affiliation{%
  \institution{Alibaba Group}
  \city{Hangzhou}
  \country{China}
}
\email{huning.hu@alibaba-inc.com}

\renewcommand{\shortauthors}{Shaochuan Lin et al.}

\begin{abstract}
  Online Food Recommendation Service (OFRS) has remarkable spatiotemporal characteristics and the advantage of being able to conveniently satisfy users' needs in a timely manner. There have been a variety of studies that have begun to explore its spatiotemporal properties, but a comprehensive and in-depth analysis of the OFRS spatiotemporal features is yet to be conducted. Therefore, this paper studies the OFRS based on three questions: how spatiotemporal features play a role; why self-attention cannot be used to model the spatiotemporal sequences of OFRS; and how to combine spatiotemporal features to improve the efficiency of OFRS. Firstly, through experimental analysis, we systemically extracted the spatiotemporal features of OFRS, identified the most valuable features and designed an effective combination method. Secondly, we conducted a detailed analysis of the spatiotemporal sequences, which revealed the shortcomings of self-attention in OFRS, and proposed a more optimized spatiotemporal sequence method for replacing self-attention. In addition, we also designed a Dynamic Context Adaptation Model to further improve the efficiency and performance of OFRS. Through the offline experiments on two large datasets and online experiments for a week, the feasibility and superiority of our model were proven.
\end{abstract}





\maketitle

\begin{table}
\small
  \caption{Ablation Study of Spatiotemporal Features. `F', `H', `T', `W', `G', `C', `A', `Ct', `Us', `It', `Sq', `BiN' are the abbreviation of Field, Hour, Time period, Week, Geohash, City ID, AOI ID, Context, User, Item, Sequence and Bias Net respectively. $\triangle$ represents the relative improvement.}
  \vskip -10pt
  \newcommand{\tabincell}[2]{\begin{tabular}{@{}#1@{}}#2\end{tabular}}
\begin{tabular}{c|c|c|c|c|c|c|c|c|c|c}\toprule
    \hline
    % Model & $\mathcal{D}_{1}$ & $\mathcal{D}_{2}$ & $\mathcal{D}_{3}$ \\
    F & H & T & W & G & C & A & AUC & $\triangle$(\%) & GAUC & $\triangle$(\%) \\
    \hline\hline
    % \ding{55}
    ~ & - & - & - & - & - & - & 0.6976 & - & 0.6029 & - \\
    \hline
   \multirow{12}*{Ct} & \ding{51} & \ding{51} & \ding{51} & \ding{51} & \ding{51} & \ding{51} & 0.6977 & 0.01 & 0.6038 & 0.15 \\
    \cline{2-11}
     & \ding{51} & - & - & - & - & -  & 0.7034 & 0.83 & 0.6100 & 1.18 \\
    \cline{2-11}
    & - & \ding{51} & - & - & - & - & 0.6751 & -3.23  & 0.5875 & -2.55 \\
    \cline{2-11}
    & - & - & \ding{51} & - & - & -  & 0.7050 & 1.06  & 0.6131 & 1.69 \\
    \cline{2-11}
    & - & - & - & \ding{51} & - & -  & 0.6980 & 0.06 & 0.6023 & -0.10 \\
    \cline{2-11}
    & - & - & - & - & \ding{51} &-  & 0.6942 & -0.49 & 0.5990 & -0.65 \\
    \cline{2-11}
    & - & - & - & - & - & \ding{51}  & 0.6887 & -1.28 & 0.5929 & -1.66 \\
    \cline{2-11}
    &\ding{51} & - & \ding{51} & - & - & -  & 0.6991 & 0.22 & 0.6050 & 0.35 \\
    \cline{2-11}
    & \ding{51} & - & - & \ding{51} & - & -  & 0.7052 & 1.09 & 0.6107 & 1.29  \\
    \cline{2-11}
    & \ding{51} & - & \ding{51} & \ding{51} & - & -  & 0.6985 & 0.13 & 0.6046 & 0.28 \\
    \cline{2-11}
    & - & - & \ding{51} & \ding{51} & - & -  & 0.7072 & 1.34 & 0.6130 & 1.68  \\
    \hline
    \hline
    \multirow{3}*{\tabincell{c}{Ct\\(w/o\\BiN)}} & \ding{51} & - & - & - & - &- & 0.6984 & 0.11 & 0.6107 & 1.29 \\
    \cline{2-11}
    ~& \ding{51} & - & - & \ding{51} & - &- & 0.6998 & 0.32 & 0.6105 & 1.26\\
    \cline{2-11}
    ~& - & - & \ding{51} & \ding{51} & - &- & 0.7012 & 0.52 & 0.6107 & 1.29\\
    % \hline
    \hline
    \hline
    \multirow{7}*{Us} & \ding{51} & \ding{51} & \ding{51} & \ding{51} & \ding{51} & \ding{51} & 0.7073 & 1.39 & 0.6132 & 1.71 \\
    \cline{2-11}
    & \ding{51} & - & - & - & - & -  & 0.7038 & 0.89 & 0.6107 & 1.30 \\
    \cline{2-11}
    & - & \ding{51} & - & - & - & - & 0.6938 & -0.55  & 0.5994 & -0.58 \\
    \cline{2-11}
    & - & - & \ding{51} & - & - & -  & 0.7047 & 1.02  & 0.6116 & 1.45 \\
    \cline{2-11}
    & - & - & - & \ding{51} & - & -  & 0.7003 & 0.39 & 0.6059 & 0.49 \\
    \cline{2-11}
    & - & - & - & - & \ding{51} &-  & 0.7020 & 0.63 & 0.6079 & 0.82 \\
    \cline{2-11}
    & - & - & - & - & - & \ding{51}  & 0.7025 & 0.70 & 0.6076 & 0.78 \\
    \hline
    \hline
    \multirow{7}*{It} & \ding{51} & \ding{51} & \ding{51} & \ding{51} & \ding{51} & \ding{51} & 0.7041 & 0.93 & 0.6104 & 1.24 \\
    \cline{2-11}
    & \ding{51} & - & - & - & - & -  & 0.7020 & 0.64 & 0.6082 & 0.88 \\
    \cline{2-11}
    & - & \ding{51} & - & - & - & - & 0.7061 & 1.22  & 0.6129 & 1.66 \\
    \cline{2-11}
    & - & - & \ding{51} & - & - & -  & 0.7054 & 1.12  & 0.6128 & 1.64 \\
    \cline{2-11}
    & - & - & - & \ding{51} & - & -  & 0.7053 & 1.11 & 0.6118 & 1.47 \\
    \cline{2-11}
    & - & - & - & - & \ding{51} &-  & 0.7043 & 0.96 & 0.6092 & 1.05 \\
    \cline{2-11}
    & - & - & - & - & - & \ding{51}  & 0.6995 & -1.28 & 0.6053 & 0.40 \\
    \hline
    \hline
    \multirow{7}*{Sq} & \ding{51} & \ding{51} & \ding{51} & \ding{51} & \ding{51} & \ding{51} & 0.7068 & 1.31 & 0.6119 & 1.50 \\
    \cline{2-11}
    & \ding{51} & - & - & - & - & -  & 0.7041 & 0.93 & 0.6113 & 1.40 \\
    \cline{2-11}
    & - & \ding{51} & - & - & - & - & 0.7030 & 0.78  & 0.6083 & 0.90 \\
    \cline{2-11}
    & - & - & \ding{51} & - & - & -  & 0.7037 & 0.87  & 0.6098 & 1.15 \\
    \cline{2-11}
    & - & - & - & \ding{51} & - & -  & 0.7050 & 1.06 & 0.6112 & 1.38 \\
    \cline{2-11}
    & - & - & - & - & \ding{51} &-  & 0.7027 & 0.73 & 0.6085 & 0.93 \\
    \cline{2-11}
    & - & - & - & - & - & \ding{51}  & 0.7045 & 0.99 & 0.6099 & 1.16 \\
    \hline
    \hline
    Ct & - & - & \ding{51} & \ding{51} & - & -  & \multirow{2}*{0.7061} & \multirow{2}*{1.22} & \multirow{2}*{0.6117} & \multirow{2}*{1.46} \\
    \cline{1-7}
    Us & \ding{51} & \ding{51} & \ding{51} & \ding{51} & \ding{51} & \ding{51} &&  &  &  \\
    \hline
    \hline
    Ct & - & - & \ding{51} & \ding{51} & - & -  & \multirow{2}*{0.7034} & \multirow{2}*{0.83} & \multirow{2}*{0.6087} & \multirow{2}*{0.96} \\
    \cline{1-7}
    It & - & \ding{51} & - & - & - & - &&& \\
    \hline
    \hline
    Ct & - & - & \ding{51} & \ding{51} & - & -  & \multirow{2}*{0.7055} & \multirow{2}*{1.13} & \multirow{2}*{0.6111} & \multirow{2}*{1.36} \\
    \cline{1-7}
    Sq & \ding{51} & \ding{51} & \ding{51} & \ding{51} & \ding{51} & \ding{51} &&& \\
    \hline
    \hline
    Ct & - & - & \ding{51} & \ding{51} & - & -  & \multirow{2}*{0.7074} & \multirow{2}*{1.40} & \multirow{2}*{0.6131} & \multirow{2}*{1.69} \\
    \cline{1-7}
    Sq & - & - & - & - & \ding{51} & - &&& \\
    \hline
    \hline
    Ct & - & - & \ding{51} & \ding{51} & - & -  & \multirow{4}*{0.7067} & \multirow{4}*{1.30} & \multirow{4}*{0.6134} & \multirow{4}*{1.74} \\
    \cline{1-7}
    Sq & - & - & - & - & \ding{51} & - &&& \\
    \cline{1-7}
    User & - & - & - & \ding{51} & - & - &&& \\
    \cline{1-7}
    Item & \ding{51} & - & - & - & - & - &&& \\
    \hline
    \hline
    Ct & - & - & \ding{51} & \ding{51} & - & -  & \multirow{4}*{\textbf{0.7077}} & \multirow{4}*{\textbf{1.45}} & \multirow{4}*{\textbf{0.6143}} & \multirow{4}*{\textbf{1.89}} \\
    \cline{1-7}
    Sq & - & - & - & - & \ding{51} & - &&& \\
    \cline{1-7}
    User & - & - & - & \ding{51} & - & - &&& \\
    \cline{1-7}
    Item & \ding{51} & \ding{51} & \ding{51} & \ding{51} & \ding{51} & \ding{51} &&& \\
    \hline
    % \hline
    % Ct & - & - & \ding{51} & \ding{51} & - & -  & \multirow{3}*{} & \multirow{3}*{} & \multirow{3}*{} & \multirow{3}*{} \\
    % \cline{1-7}
    % Sq & - & - & \ding{51} & \ding{51} & - & - &&& \\
    % \cline{1-7}
    % It & - & - & \ding{51} & \ding{51} & - & - &&& \\
    \hline
\end{tabular}
  \label{table:st_features}
  % \vskip -5pt
\end{table}

\section{Introduction}
Online Food Recommendation Service (OFRS) is a powerful and efficient service that provides users with the ability to swiftly recommend food in a designated region and time period, thus satisfying their needs promptly. In recent years, there have been a number of studies devoted to exploring OFRS. For example, TRISAN\cite{TRISAN} models the online food recommendation service using the triangular relationship consisting of user location, merchant location, and time, and achieved good results. STPIL\cite{STPIL} has conducted an in-depth exploration of the periodic characteristics of user behavior. StEN\cite{StEN} models the user's spatiotemporal characteristics within OFRS from the perspective of static and dynamic spatiotemporal features. BASM\cite{BASM} enhances the fitting ability of the model under complex spatiotemporal distribution from the perspective of dynamic parameter modeling.

In spite of the earlier investigations into the spatiotemporal models of OFRS, a comprehensive understanding of their spatiotemporal characteristics has yet to be achieved. Specifically, which spatiotemporal features are essential in the online food recommendation service? Moreover, in contrast to other recommendation scenarios, user sequence modeling in this context has not yet adopted Self-attention mechanism, which is known to be highly effective. The underlying reason behind this remains unknown. Consequently, further research is needed to investigate how to leverage spatiotemporal features to develop an efficient spatiotemporal model. In the following section, we discuss the three topics in greater detail.

In the online food recommendation service, we can generally extract the context features in the ranking stage, such as the Hour, Time period (including breakfast, lunch, afternoon tea, dinner, and supper), Geohash of the user's current location, $etc$. We first proposed a straightforword Baseline Model to capture the varios spatiotemporal characteristics by an innovative Bias Net. Through an extensive series of elimination experiments on temporal and spatial features, we discovered that `Week' of time are much more influential than previously thought, which differs from our initial intuition that the Time period would be the most influential factor. What's more, we also found that richer spatiotemporal features do not necessarily show significant improvement in model performance, which is mostly ignored in most studies. In addition, accurate selection of specific spatiotemporal features can quickly improve the model performance by one level without the need for complex models and operations, which has not been given enough attention in most literatures. Furthermore, the combination of different spatiotemporal features into their respective feature fields will encounter different performance, so this should be fully considered in the subsequent model construction.

% Through established ablation experiments, the role of temporal features was found to be more pronounced than spatial features, particularly with regard to the influence of the time period on user selection. Moreover, incorporating spatiotemporal features as a distinct feature field would result in a negative impact on the model. It is essential to consider the integration of spatiotemporal features with other features in order to enhance the performance of the model.

\begin{figure}[tbp]
\vskip -10pt
\centerline{\includegraphics[width=.8\columnwidth]{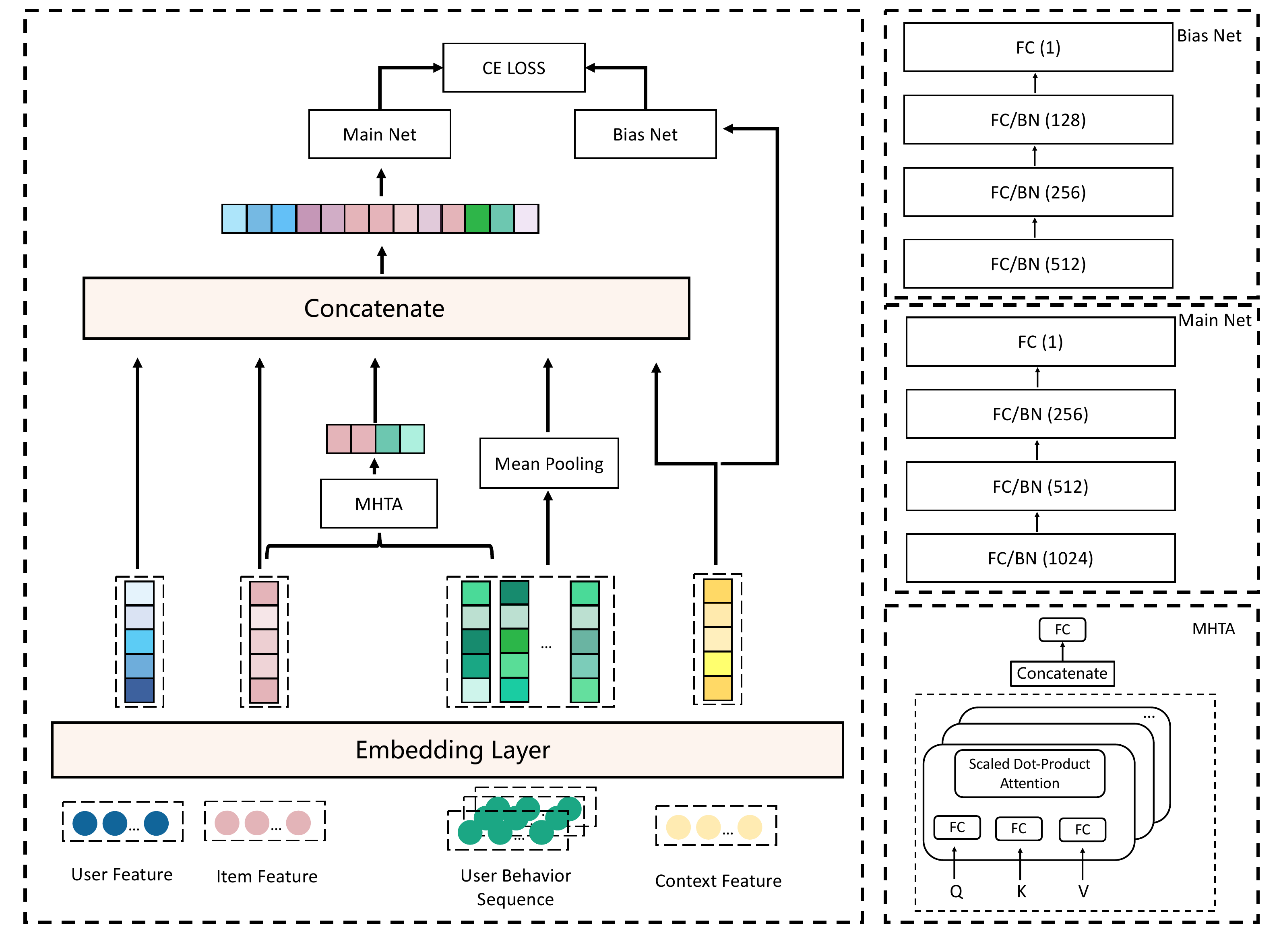}}
\vskip -10pt
\caption{An illustration of Baseline Model for OFRS.}
\vskip -18pt
\label{fig:base_model}
% \vskip -5pt
\end{figure}

The current online food recommendation services rarely utilize the self-attention mechanism to model user sequences, yet this mechanism has become popularly incorporated into other fields \cite{DIEN,bst,STAN,TiSAS}. To evaluate its effectiveness in OFRS, we experimented with the self-attention module, but the outcome was not satisfactory. Subsequent analysis of the low performance was conducted and it was hypothesized that the utilization of self-attention was heavily reliant on the relationship between user spatiotemporal sequences, which could potentially introduce substantial temporal and spatial sequence distortion as a result of users’ varying preferences and interests across different temporal and spatial contexts. In current works, target attention is usually used for sequence modeling, while the time sequence information is invariably omitted. To tackle this, position encoding can be used to compensate for the lost sequence order, yet this is not practicable in food recommendation systems. Hence, we propose a method of calculating the time difference and inserting it directly into the sequence as a feature. Experimental results demonstrate that this method is both simple and effective.

In aiming to further enhance the utility of spatiotemporal features in online food recommendation services, we propose a Dynamic Context Adaptation Model based on the conclusions mentioned above. In our model, we exploit spatiotemporal features to dynamically adjust the integration of different feature fields without time-consuming feature selection.  The offline results on two large industrial datasets and the one-week online results have both demonstrated the effectiveness of our model.

% We then further proposed a personal time period constraint module to further improve the model performance.

% Subsequent experiments on two massive industrial datasets demonstrate the effectiveness of our proposed model, and we deploy it to a real online food recommendation system. After careful observation over a week, we observe that our model delivers excellent results online.

% presenting that spatiotemporal features play a role in recommendation and proposing a strong baseline model for OFRS.

% Overall, the contributions of this paper can be summarized as:
% \begin{itemize}
%     \item  presenting the significance of spatiotemporal features and proposing a robust baseline model for Online Food Recommendation Systems(OFRS).
%     \item elucidating why the Self-attention model performs poorly in food recommendation and proposing a method to calculate time difference and incorporate it into the sequence as a feature.
%     \item proposing an institute spatiotemporal perception model based on spatiotemporal features for online recommendation and demonstrating its effectiveness in online scenarios through experiments.
% \end{itemize}

\section{Influence of Spatiotemporal Features}
In this section, we investigate the impact of spatiotemporal features in OFRS and present a generic Baseline Model based on the spatiotemporal characteristics of OFRS.

% In the online food recommendation service, the general model goes through at least two stages of recall-ranking. In the first phase of recall, and based on the user's current geographic location, a large number of merchants that are not within a certain range of the user or do not meet the delivery distance requirements are filtered. Therefore, when recommending products in the sorting stage, the candidate products must have certain geographical requirements. Therefore, when exploring the time and space of the online food recommendation service for users, the more powerful feature is actually the time feature.
\begin{table}
\small
\vskip -10pt
  \caption{Sequence Modeling.}
  \vskip -10pt
  \begin{tabular}{c|c|c}
    \hline
    % Model & $\mathcal{D}_{1}$ & $\mathcal{D}_{2}$ & $\mathcal{D}_{3}$ \\
    Method & AUC &  GAUC  \\
    \hline
    GRU & 0.7051 & 0.6113  \\
    \hline
    Self-Attention & 0.6922 & 0.5972 \\
    \hline
    Self-Attention+Target Attention & 0.6923 & 0.5978  \\
    \hline
    Target Attention & 0.7074 & 0.6131  \\
    % todo
    \hline
    Target Attention+Position Encoding & 0.7033 & 0.6089  \\
    \hline
    Target Attention+Time Difference & 0.7081 & 0.6150  \\
    % todo
    
    \hline
    
\end{tabular}
  \label{table:attention}
  \vskip -10pt
\end{table}

\subsection{Baseline Model}
Fig.~\ref{fig:base_model} illustrates the Baseline Model structure for OFRS, wherein $x= (u, i, b, c) \in \mathcal{X}$ is the input data and $y \in \mathcal{Y}$ is the click label. The features of users, items, behavior sequences, and contexts are represented as $u$, $i$, $b$, and $c$, respectively. By leveraging an embedding layer, user embedding $e(u)$, item embedding $e(i)$, behavior sequence emebdding $e(b)$, and context embedding $e(c)$ can be obtained. Through Multi-head Target Attention (MHTA), user interests are represented as user interest representation,
\begin{align}
    S & = MHTA(e(i), e(b))  =Concat(head_{0},head_{1},..., head_{h-1}), \\
        & where\ head_i = Attention (e(i)W^{Q}, e(b)W^{K}, e(b)W^{V}), \\
        & Attention(Q, K, V) = softmax(\frac{QK^{T}}{\sqrt{d}})V
\end{align}
where projection matrices $W^{Q} \in \mathbb{R}^{d_i \times d}, W^{K},W^{V} \in \mathbb{R}^{d \times d}$. $d_i$ and $d$ are the dimension of $e(i)$ and $e(b)$, respectively. We then concatenate all the above embedding through, 
\begin{equation}
	\begin{split}
	o &= MLP_{M}(Concat(e(u),e(i), S, MP(e(b)), e(c))) \\
	        & + MLP_{B}(e(c))
	\end{split}
\end{equation}
where $MLP_{M}$ and $MLP_{B}$ are the Main Net and Bias Net respectively. $MP$ represents Mean Pooling. Detail structure can be seen in Fig.~\ref{fig:base_model}. It is necessary to emphasize that there exist distinct variations in Click-Through Rate (CTR) when it comes to time and location of OFRS. Thus, as a result, our Baseline Model was designed with temporal and spatial features as bias signal, allowing the model to gain a clearer recognition of the different temporal and spatial characteristics of OFRS. Table~\ref{table:st_features} `Ct(w/o BiN)' provides the ablation study of Bias Net.
Finally, we optimize the whole network by Cross-entropy Loss (CE Loss): $\mathcal{L} =  -\frac{1}{N} \sum ylog(p(o)) + (1-y)log(1-p(o))$, where $p(o)$ is the output obtained following sigmoid activation. $N$ is the mini-batch size.

% (CE Loss):
% \begin{equation}
%   \mathcal{L} =  -\frac{1}{N} \sum ylog(p(o)) + (1-y)log(1-p(o))
% \end{equation}
% Where $p(o)$ is the output obtained following sigmoid activation. $N$ is the mini-batch size.

\subsection{ Ablation Study of Spatiotemporal Features}
In OFRS, we can generally extract various context features, including the Hour, Time period ($i.e.$, breakfast, lunch, afternoon tea, dinner and supper), Week of the user's time, Geohash of the user's current location, City ID of the user's current location, and AOI ID (AOI ID can identify a specific community, office park, school, hospital, $etc$.). An ablation study was conducted with our Baseline Model on an industrial THEME dataset (Detail can be seen in Table~\ref{table:dataset}) to further analyze the effect of each feature. The results presented in Table~\ref{table:st_features} indicate that Week of the user's time is the most fitting feature, bringing positive returns no matter how it is combined, as is evidenced by the experiment of selecting `W' with `Ct`/`Us'/`It'/`Sq'. The experiment results also indicate that there is no consistent correlation between an increase in spatiotemporal features and improved performance (Table~\ref{table:st_features}, row 3), which has been overlooked in earlier spatiotemporal modelling studies. Additionally, precise selection of certain spatiotemporal features can rapidly elevate the model performance by one level without the demand for sophisticated models and operations, which has been largely neglected in many research papers. Moreover, profound effect was achieved as the AUC score was increased from 0.6976 to 0.7077\footnote{0.1\% increase in AUC is already obvious in the industrial dataset.}(Table~\ref{table:st_features}, last row) after manual selection of the spatiotemporal features.

% public dataset\footnote{https://tianchi.aliyun.com/dataset/131047} 
% industrial dataset (Detail can be seen in Table~\ref{table:dataset})
% \footnote{0.1\% increase in AUC is already obvious in the industrial dataset.} 

% Table~\ref{table:st_features} indicates that different spatiotemporal features have different contributions to model performance. Experiments also proved that it is not necessary for the model to have more spatiotemporal features to have better performance, which is not mentioned in many related studies discussing spatiotemporal modeling. In addition, accurate selection of specific spatiotemporal features can quickly improve the model performance by one level without the need for complex models and operations, which has not been given enough attention in most literatures.  Furthermore, model effect can quickly go from AUC 0.6976 to 0.7074\footnote{0.1\% increase in AUC is already obvious in the industrial dataset.} after manual precisely selecting spatiotemporal features.

% How to Incorporate the above Spatiotemporal features

% \begin{table}
%   \caption{Incorporate Spatiotemporal Features.}
%   \input{./table/feature_add_col.tex}
%   \label{table:feature_add_col}
% \end{table}

\section{Sequence Modeling for OFRS}
In this section, the use of spatiotemporal sequences has been explored. Some studies\cite{DIEN,bst,STAN,TiSAS} have adopted self-attention to process sequences and proved its effectiveness in practical environments. However, we rarely see the direct application of self-attention in online food recommendation systems. Therefore, we conducted experiments and the results are shown in Table~\ref{table:attention}. We compared the performance of GRU (Gated Recurrent Unit) and self-attention in extracting sequence information, and found that GRU outperformed self-attention since it can better capture time order information. Nevertheless, self-attention has better parallelism, leading to faster running speed, making it a better choice for recommendation systems with strict running time requirements. 
Moreover, Table~\ref{table:attention} shows that target attention is more suitable than self-attention+target attention in OFRS. The latter amplifies the differences between spatiotemporal sequences, thus potentially raising noise levels in the sequence, a situation exacerbated by the rapid changes in user interests over time and space ($e.g.$, between breakfast and dinner).
We then employed position encoding to reinforce temporal ordering in the attention mechanism. However, the experiment results revealed our modification to have limited improvement, so we proposed an effective alternative. We incorporated the time difference between the sequence time and the target time as a feature in the sequence so that the model could learn the temporal orderliness. The experimental results in Table~\ref{table:attention} show that this method to be effective, achieving satisfactory performance improvement.

\begin{figure}[tbp]
\vskip -10pt
\centerline{\includegraphics[width=0.6\columnwidth]{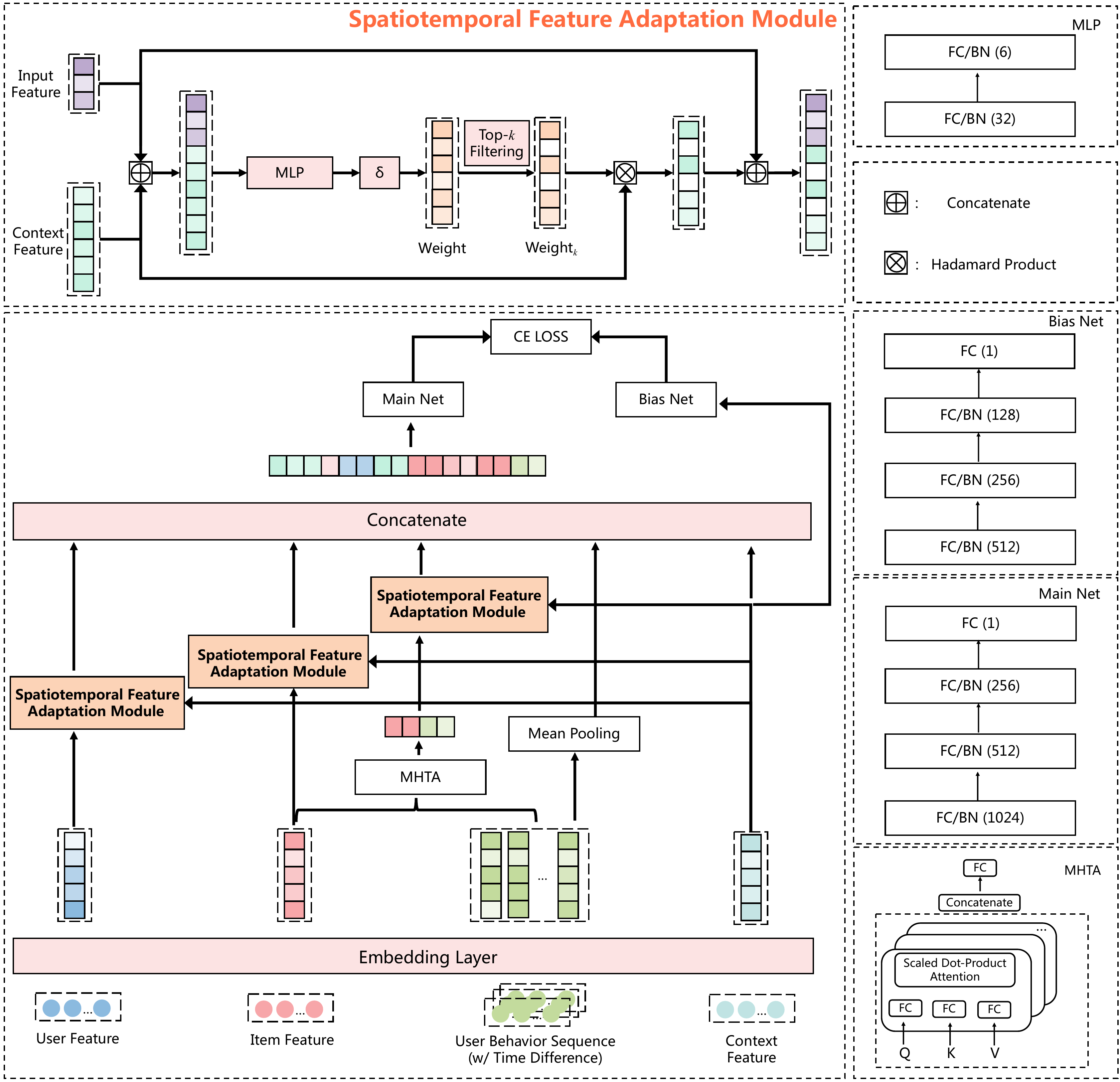}}
\vskip -10pt
\caption{The overview of Dynamic Context Adaptation Model (DCAM) structure.}
\label{fig:DCAM}
\vskip -15pt
\end{figure}

\begin{table*}
\small
  \caption{Performances of different recommendation algorithms. The best are highlighted in bold.}
  \vskip -10pt
  \begin{tabular}{c|c|c|c|c|c|c|c|c|c|c|c|c}
  \hline
    \multirow{2}*{Methods} & \multicolumn{6}{c|}{THEME} & \multicolumn{6}{c}{SHOP} \\
    \cline{2-13} 
    ~&AUC&GAUC&Logloss&NDCG1&NDCG5&NDCG10 
     % &AUC&GAUC&Logloss&NDCG1&NDCG5&NDCG10
     &AUC&GAUC&Logloss&NDCG1&NDCG5&NDCG10 \\
  \hline
    Wide\&Deep\cite{wide_deep}  
    & 0.6833 & 0.5904 & 0.0466 & 0.3348 & 0.5865 & 0.6361 & 0.7341 & 0.6219 & 0.0466 & 0.2127 & 0.4318 & 0.4819 \\
    \hline
    % BASE & 0.8002 & 0.7086 & 0.4392 & 0.2874 & 0.4926 & 0.5383 
    % & ~ & ~ & ~ & ~ & ~ & ~ \\
    DIN\cite{DIN} 
    & 0.6885 & 0.5943 & 0.0469 & 0.3375 & 0.5901 & 0.6390 & 0.7888 & 0.6840 & 0.0445 & 0.2616 & 0.4689 & 0.5165  \\
    % \hline
    % DIEN\cite{DIEN} & ~ & ~ & ~ & ~ & ~ & ~
    % & ~ & ~ & ~ & ~ & ~ & ~ \\
    \hline
    StEN\cite{StEN} 
    & 0.6909 & 0.5965 & 0.0473 & 0.3373 & 0.5889 & 0.6381 & 0.7898 & 0.6883 & 0.0444 & 0.2628 & 0.4705 & 0.5180 \\
    \hline
    BASM \cite{BASM} 
    & 0.6720 & 0.5831 & 0.0524 & 0.3417 & 0.5934 & 0.6418 & 0.7928 & 0.6956 & 0.0443 & 0.2804 & 0.4865 & 0.5324  \\
    % \hline
    % TiSASRec \cite{TiSAS} & ~ & ~ & ~ & ~ & ~ & ~ & ~ & ~ & ~ & ~ & ~ & ~\\
    % \hline
    % TimelyRec \cite{Timely} & ~ & ~ & ~ & ~ & ~ & ~ & ~ & ~ & ~ & ~ & ~ & ~\\
    % % \hline
    % % TRISAN  & ~ & ~ & ~ & ~ & ~ & ~ & ~ & ~ & ~ & ~ & ~ & ~\\
    % \hline
    % STAN\cite{STAN} & ~ & ~ & ~ & ~ & ~ & ~ & ~ & ~ & ~ & ~ & ~ & ~\\
    \hline\hline
    % w/ StFAM & 0.7866 & 0.6847 & 0.0445 & 0.2607 & 0.4683 & 0.5160 & 0.6982 & 0.6034 & 0.0446 & 0.3430 & 0.5958 & 0.6432\\
    % \hline
    % w/ PTCM & 0.7936 & 0.7001 & 0.0442 & 0.2782 & 0.4848 & 0.5309 & 0.6940 & 0.5996 & 0.0460 & 0.3411 & 0.5927 & 0.6408\\
    % \hline
    DCAM  & \textbf{0.7093} & \textbf{0.6152} & \textbf{0.0438} & \textbf{0.3594} & \textbf{0.6079} & \textbf{0.6535} & \textbf{0.7940} & \textbf{0.7003} & \textbf{0.0442} & \textbf{0.2843} & \textbf{0.4873} & \textbf{0.5333} \\
    % \hline
    % Improvement & ~ & ~ & ~ & ~ & ~ & ~ & ~ & ~ & ~ & ~ & ~ & ~\\
  \hline
\end{tabular}

  \label{table:sota}
  \vskip -10pt
\end{table*}

\section{Dynamic Context Adaptation Model}
By combining the above spatiotemporal sequence methods and the manual configuration of spatiotemporal feature, we can easily improve the model performance. However, due to the time-consuming and labor-intensive nature of this configuration method, we further proposed an efficient and simple Dynamic Context Adaptation Model(DCAM) to not only improve the model performance, but also reduce the difficulty of manual configuration of spatiotemporal features. As illustrated in Figure 2, we designed a Spatiotemporal Feature Adaptation Module (StFAM) which can be implemented by the following formula,
\begin{gather}
    \label{eq:weight}
    Weight = \delta(MLP(Concat(e(c), e(input)))) \\ 
    \label{eq:weightk}
    Weight_k = Top-k\ Filtering(Weight)  
\end{gather}
where $\delta$ is the sigmoid activation function. The final spatiotemporal feature adaptation output can be calculated from the $input$ features (User Feature, Item Feature and target-attention output). As illustrated in Fig.\ref{fig:DCAM}, the $MLP$ structure contains a last layer of dimension 6, corresponding to the 6 spatiotemporal features. To counteract the distortion of noise, the interaction $Weight$ obtained by Eq.~\ref{eq:weight} is filtered using parameter $k$, and the filtered $Weight_k$ is then utilized to yield the desired output,
\begin{gather}
    A_c = Reshape_w(Weight_k) \odot Reshape_c(e(c)) \\
    A_{output} = Concat(e(input), A_c)
\end{gather}
where the tensor $Weight_k$ is reshaped as a shape of $[B, 6, 1]$ by $Reshape_w$, and the tensor $e(c)$ is reshaped as a shape of $[B, 6, d_c]$ by $Reshape_w$. $d_c$ is the dimension of each spatiotemporal feature embedding and $\odot$ is the the Hadamard product.

% The manually-defined time period features, such as indicating which time period is breakfast and which is dinner, significantly impede users' personalization of the time period, thus resulting in a substantially restricted user experience. In order to enhance the temporal sensitivity of the model to items, we introduce a Personal Time period Constraint Module (PTCM), enabling the model to have a unique personalized time period for different users and different items.

% In order to enhance the temporal sensitivity of the model to items and allow for improved personalization of the time period, we propose a Personal Time period Constraint Module (PTCM). This module enables the model to have a unique personalized time period for different users and different items, thus resulting in an enhanced user experience,

% \begin{equation}
%     \begin{aligned}
%         o_{tp} &= MLP_{tp}(Concat(A_u, A_i, A_S)) \\
%   \mathcal{L}_{tp} &=  -\frac{1}{N} \sum (\hat{y_{tp}})log(p(o_{tp}))
%     \end{aligned}
% \end{equation}
% where $p(o_{tp})$ is the output obtained following softmax activation. $\hat{y_{tp}}$ is the label of time period, which serves as the ground truth before the $n$ iterations have elapsed. After that, it is substituted by the self-supervised label that is equivalent to $p(o_{tp})$, thereby facilitating the optimization of the system and the concomitant enhancement of the results. The predefined Time period feature for breakfast/dinner impede users' personalization of the time period, significantly compromising their user experience.

\begin{figure}[hbp]
% \vskip -15pt
% \centerline{\includegraphics[width=0.3\columnwidth]{./fig/topk.pdf}}
% \vskip -15pt
% \caption{Parameter analysis of $k$}
% \label{fig:topk}
% \vskip -20pt
\vskip -10pt
\subfigure[Parameter analysis]{ \label{fig:topk} 
\includegraphics[width=0.45\columnwidth]{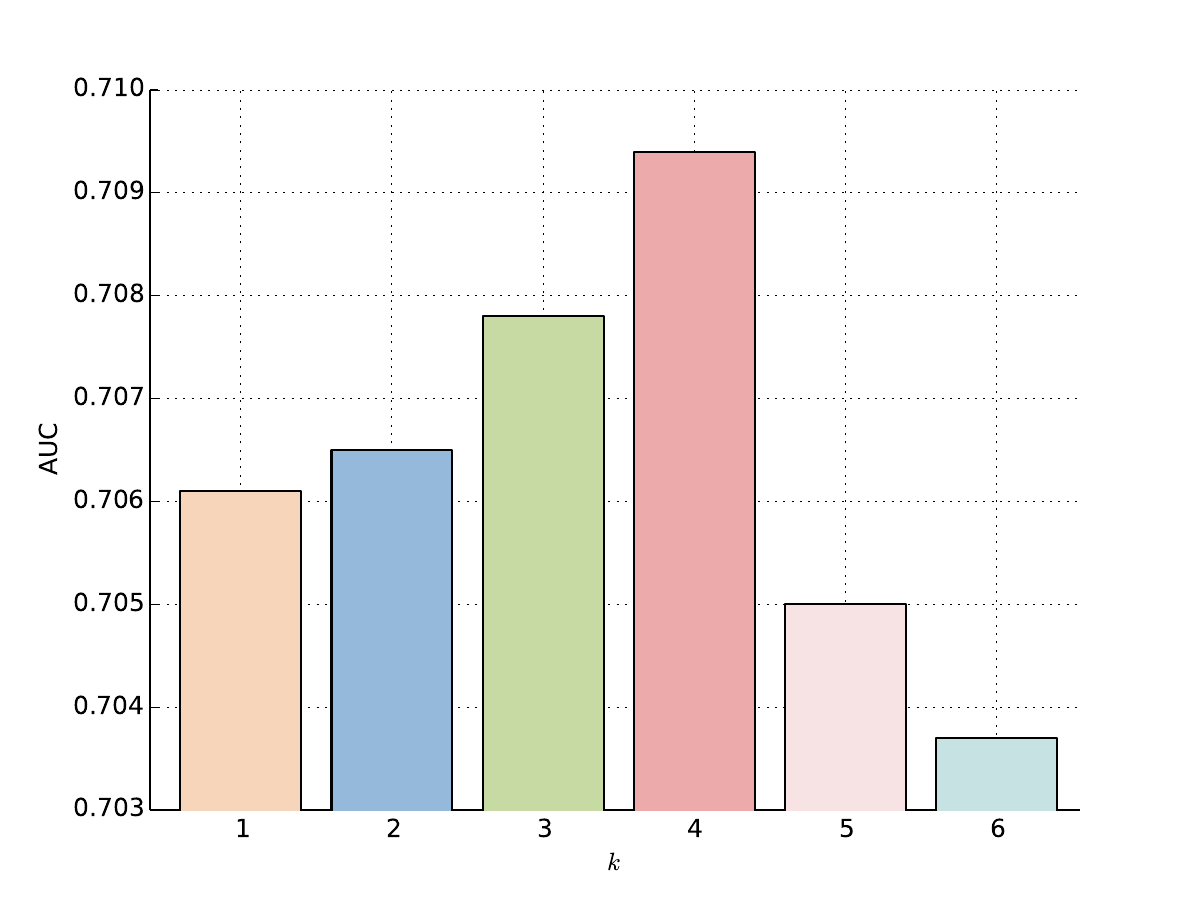}
} 
\subfigure[Online performance]{ \label{fig:online} 
\includegraphics[width=0.45\columnwidth]{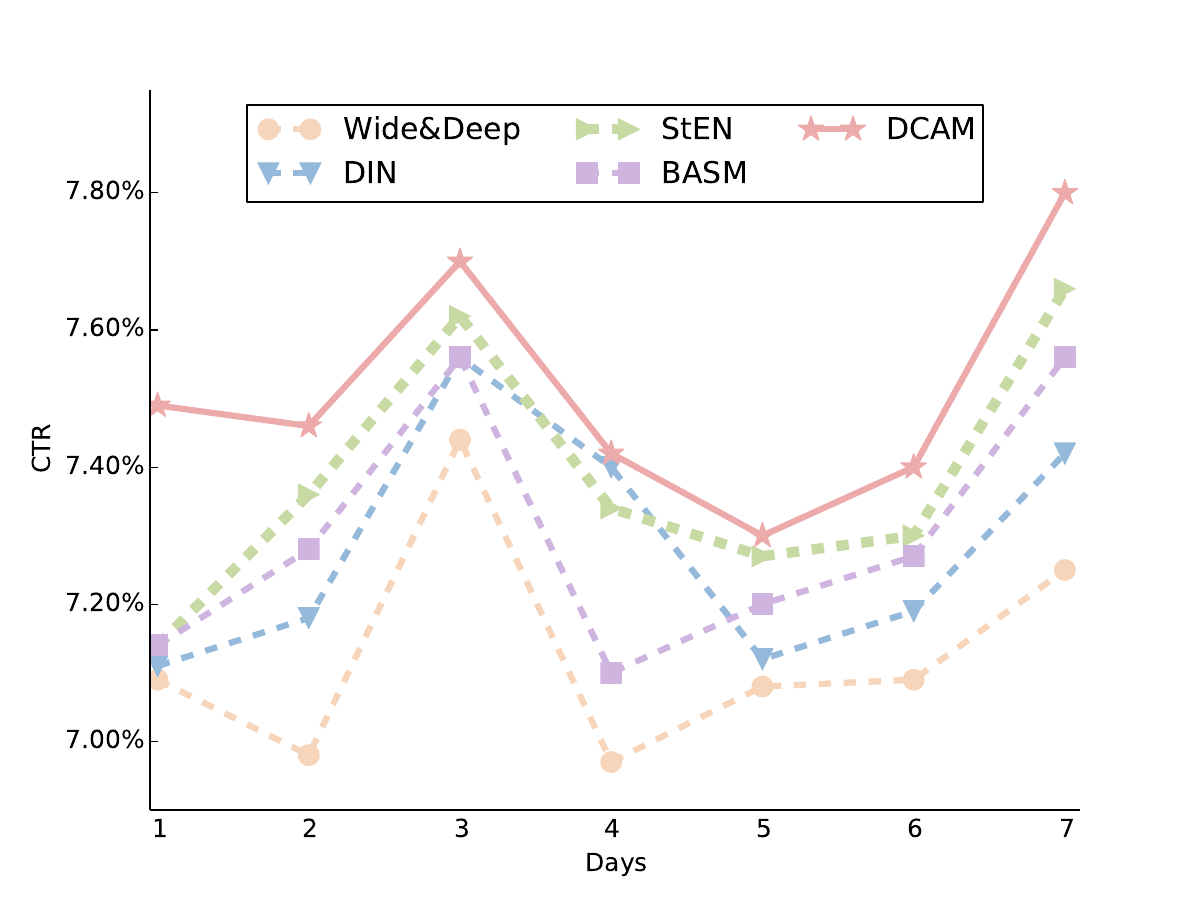}
} 
\vskip -10pt
\caption{ (a) Parameter analysis of $k$. (b) Online performance in one week of February 2023. } 
\vskip -10pt
\end{figure}

\section{Experimental Results}
% In this section, we conduct experiments to answer the following research questions:
% \begin{itemize}
%     \item \textbf{RQ1} How does DCAM perform compared with state-of-the-art(SOTA) methods?
%     \item \textbf{RQ2} Do components in DCAM necessarily contribute to performance?
%     \item \textbf{RQ3} How do DCAM and other methods perform online?
% \end{itemize}

\subsection{Datasets and Evaluation Metrics}
% \subsubsection{Datasets} 
We conduct experiments on two large datasets from real-world online food recommendation platforms. 1) THEME dataset, taking logs from Nov. 1 to Jan. 14 as training data and Jan. 15 as test data, with over 300 features. 2) SHOP dataset, comprising continuous 7-day data for training and 1 consecutive day data for testing, with more than 8.8 billion samples in total. Detailed information can be seen in  Table~\ref{table:dataset}. In order to comprehensively evaluate the offline effect of the model, we use several commonly used indicators in the recommendation system to measure, including Area Under Curve(AUC), Group Area Under Curve(GAUC), Logloss, and Normalized Discounted Cumulative Gain(NDCG).

% \subsubsection{Metrics} 

\begin{table}
\small
  \caption{Statistics of datasets.}
  \vskip -10pt
  \begin{tabular}{c|c|c|c|c}
    \hline
    % Model & $\mathcal{D}_{1}$ & $\mathcal{D}_{2}$ & $\mathcal{D}_{3}$ \\
    Dataset & \#Users & \#Items & \#Samples & \#Features \\
    \hline
    THEME & 39950940 & 14161762 & 1104609412  & 316 \\
    \hline
    SHOP & 74917907 & 2660149 & 8126820568 & 170 \\
    
    \hline
\end{tabular}
  \label{table:dataset}
  \vskip -15pt
\end{table}

\subsection{Parameter settings and benchmarks}
% \subsubsection{Parameter settings} 
In all experiments, mini-batch $N$ of this paper are set to 1024, and AdaGray\cite{adagray} is adopted as the optimizer. In order to enhance the performance of the model, Warm-up\cite{warmup} mechanism has been adopted for all models, with the learning rate increasing from 0.001 to 0.015 from 0 to 1 million steps. $k$ is set to 4 by grid search. The experimental environment is using Python 2.7 and TensorFlow 1.14. Meanwhile, all experiments are conducted on our self-developed platform, with 40 parameter servers and 400 workers without relying on GPU. We evaluate our proposed approach against several state-of-the-art benchmark methods, including Wide\&Deep\cite{wide_deep}, DIN\cite{DIN}, StEN\cite{StEN} and BASM\cite{BASM}.

% \subsubsection{Benchmarks} 
% \textbf{Wide\&Deep\cite{wide_deep}} divides features into two parts: a deep side for learning the generalization ability of the model and a wide side for learning the memorization ability.
% \textbf{DIN\cite{DIN}} introduces the attention mechanism to more accurately capture user interests without compromising the controllable model parameters and computational complexity.
% \textbf{StEN\cite{StEN}} segregates user temporal and spatial features into static attributes and dynamic interests, and models user temporal and spatial characteristics from both static and dynamic perspectives. 
% \textbf{BASM\cite{BASM}} believes that using a set of fixed parameters is difficult to fit complex data distributions, so dynamic parameter mechanisms are adopted to fit spatiotemporal data distributions.

\subsection{Offline and Online Performance}
Experiments conducted on the THEME and SHOP datasets (Table~\ref{table:sota}) revealed that DCAM outperformed the other models in terms of AUC, GAUC, and NCDG indices, while the Logloss index was lower. We also analyze the impact of $k$ on THEME dataset (Fig.~\ref{fig:topk}), which shows that $k=4$ yields the best performance.

Offline comparison experiments previously verified the superiority of the DCAM model. To further corroborate this conclusion, an A/B test was conducted on a real online food recommendation system in February 2023, deploying DCAM and 4 other models. Results in Fig.~\ref{fig:online} reveal that DCAM exhibited outstanding performance online, significantly outperforming the other models.

\section{Conclusion}

% In conclusion, the potential of Online Food Recommendation Service's spatiotemporal features has been realized through this paper. An effective method of combining spatiotemporal features was designed and tested, as was a new spatiotemporal sequence method to replace self-attention. Further, a spatiotemporal perception algorithm was also proposed. The results of the experiments proved the validity and superiority of this model. Thus, it is safe to say that this paper contains valuable insight into the efficient application of spatiotemporal features in OFRS, and further exploration of this topic is needed.

This paper investigates the value of spatiotemporal features in Online Food Recommendation Service (OFRS). Experiments demonstrate that effective selection of spatiotemporal features can produce significant improvements. Furthermore, we discovered that self-attention is not suitable in OFRS due to the possible introduction of substantial temporal and spatial sequence distortion, and hence a Dynamic Context Adaptation Model (DCAM) has been developed to further improve the performance and efficiency of OFRS. The results of the experiments validate the superiority of this approach. Therefore, this paper offers useful insight into the effective use of spatiotemporal features in OFRS, and further studies should be conducted to explore this topic more deeply.

\section*{COMPANY PORTRAIT}

Alibaba Group is a comprehensive enterprise group in China that provides a full range of Internet services, ranging from China commerce, international commerce, local consumer services, cloud computing, digital media and entertainment, $etc$.  As a part of the Alibaba Group, Ele.me is a leading online food recommendation service (OFRS) and on-demand delivery platform in China, providing services to over one hundred million users daily.

\section*{PRESENTER BIO}

Shaochuan Lin is currently employed as an Algorithm Engineer at Ele.me-Alibaba Group, where he is responsible for the homepage recommendation of the Ele.me App. He received the master degree from South China University of Technology and is currently focusing his efforts in the areas of Search Recommendation \& Content Analysis for Search and Recommendation. 

\balance 

\bibliographystyle{ACM-Reference-Format}

\bibliography{paper}


\begin{thebibliography}{12}

%%% the Web address.

\ifx \showCODEN    \undefined \def \showCODEN     #1{\unskip}     \fi
\ifx \showDOI      \undefined \def \showDOI       #1{#1}\fi
\ifx \showISBNx    \undefined \def \showISBNx     #1{\unskip}     \fi
\ifx \showISBNxiii \undefined \def \showISBNxiii  #1{\unskip}     \fi
\ifx \showISSN     \undefined \def \showISSN      #1{\unskip}     \fi
\ifx \showLCCN     \undefined \def \showLCCN      #1{\unskip}     \fi
\ifx \shownote     \undefined \def \shownote      #1{#1}          \fi
\ifx \showarticletitle \undefined \def \showarticletitle #1{#1}   \fi
\ifx \showURL      \undefined \def \showURL       {\relax}        \fi
% invisible to TeX
\providecommand\bibfield[2]{#2}
\providecommand\bibinfo[2]{#2}
\providecommand\natexlab[1]{#1}
\providecommand\showeprint[2][]{arXiv:#2}

\bibitem[Chen et~al\mbox{.}(2019)]%
        {bst}
\bibfield{author}{\bibinfo{person}{Qiwei Chen}, \bibinfo{person}{Huan Zhao},
  \bibinfo{person}{Wei Li}, \bibinfo{person}{Pipei Huang}, {and}
  \bibinfo{person}{Wenwu Ou}.} \bibinfo{year}{2019}\natexlab{}.
\newblock \showarticletitle{Behavior Sequence Transformer for E-commerce
  Recommendation in Alibaba}.
\newblock \bibinfo{journal}{\emph{CoRR}}  \bibinfo{volume}{abs/1905.06874}
  (\bibinfo{year}{2019}).
\newblock


\bibitem[Cheng et~al\mbox{.}(2016)]%
        {wide_deep}
\bibfield{author}{\bibinfo{person}{Heng{-}Tze Cheng}, \bibinfo{person}{Levent
  Koc}, \bibinfo{person}{Jeremiah Harmsen}, \bibinfo{person}{Tal Shaked},
  \bibinfo{person}{Tushar Chandra}, \bibinfo{person}{Hrishi Aradhye},
  \bibinfo{person}{Glen Anderson}, \bibinfo{person}{Greg Corrado},
  \bibinfo{person}{Wei Chai}, \bibinfo{person}{Mustafa Ispir},
  \bibinfo{person}{Rohan Anil}, \bibinfo{person}{Zakaria Haque},
  \bibinfo{person}{Lichan Hong}, \bibinfo{person}{Vihan Jain},
  \bibinfo{person}{Xiaobing Liu}, {and} \bibinfo{person}{Hemal Shah}.}
  \bibinfo{year}{2016}\natexlab{}.
\newblock \showarticletitle{Wide {\&} Deep Learning for Recommender Systems}.
  In \bibinfo{booktitle}{\emph{Proceedings of the 1st Workshop on Deep Learning
  for Recommender Systems, DLRS@RecSys 2016, Boston, MA, USA, September 15,
  2016}}, \bibfield{editor}{\bibinfo{person}{Alexandros Karatzoglou},
  \bibinfo{person}{Bal{\'{a}}zs Hidasi}, \bibinfo{person}{Domonkos Tikk},
  \bibinfo{person}{Oren~Sar Shalom}, \bibinfo{person}{Haggai Roitman},
  \bibinfo{person}{Bracha Shapira}, {and} \bibinfo{person}{Lior Rokach}}
  (Eds.). \bibinfo{publisher}{{ACM}}, \bibinfo{pages}{7--10}.
\newblock
\urldef\tempurl%
\url{https://doi.org/10.1145/2988450.2988454}
\showDOI{\tempurl}


\bibitem[Cui et~al\mbox{.}(2021)]%
        {STPIL}
\bibfield{author}{\bibinfo{person}{Qiang Cui}, \bibinfo{person}{Chenrui Zhang},
  \bibinfo{person}{Yafeng Zhang}, \bibinfo{person}{Jinpeng Wang}, {and}
  \bibinfo{person}{Mingchen Cai}.} \bibinfo{year}{2021}\natexlab{}.
\newblock \showarticletitle{{ST-PIL:} Spatial-Temporal Periodic Interest
  Learning for Next Point-of-Interest Recommendation}. In
  \bibinfo{booktitle}{\emph{{CIKM} '21: The 30th {ACM} International Conference
  on Information and Knowledge Management, Virtual Event, Queensland,
  Australia, November 1 - 5, 2021}},
  \bibfield{editor}{\bibinfo{person}{Gianluca Demartini},
  \bibinfo{person}{Guido Zuccon}, \bibinfo{person}{J.~Shane Culpepper},
  \bibinfo{person}{Zi~Huang}, {and} \bibinfo{person}{Hanghang Tong}} (Eds.).
  \bibinfo{publisher}{{ACM}}, \bibinfo{pages}{2960--2964}.
\newblock


\bibitem[Du et~al\mbox{.}(2022)]%
        {BASM}
\bibfield{author}{\bibinfo{person}{Boya Du}, \bibinfo{person}{Shaochuan Lin},
  \bibinfo{person}{Jiong Gao}, \bibinfo{person}{Xiyu Ji},
  \bibinfo{person}{Mengya Wang}, \bibinfo{person}{Taotao Zhou},
  \bibinfo{person}{Hengxu He}, \bibinfo{person}{Jia Jia}, {and}
  \bibinfo{person}{Ning Hu}.} \bibinfo{year}{2022}\natexlab{}.
\newblock \showarticletitle{{BASM:} {A} Bottom-up Adaptive Spatiotemporal Model
  for Online Food Ordering Service}.
\newblock \bibinfo{journal}{\emph{CoRR}}  \bibinfo{volume}{abs/2211.12033}
  (\bibinfo{year}{2022}).
\newblock


\bibitem[Duchi et~al\mbox{.}(2011)]%
        {adagray}
\bibfield{author}{\bibinfo{person}{John Duchi}, \bibinfo{person}{Elad Hazan},
  {and} \bibinfo{person}{Yoram Singer}.} \bibinfo{year}{2011}\natexlab{}.
\newblock \showarticletitle{Adaptive subgradient methods for online learning
  and stochastic optimization.}
\newblock \bibinfo{journal}{\emph{Journal of machine learning research}}
  \bibinfo{volume}{12}, \bibinfo{number}{7} (\bibinfo{year}{2011}).
\newblock


\bibitem[He et~al\mbox{.}(2016)]%
        {warmup}
\bibfield{author}{\bibinfo{person}{Kaiming He}, \bibinfo{person}{Xiangyu
  Zhang}, \bibinfo{person}{Shaoqing Ren}, {and} \bibinfo{person}{Jian Sun}.}
  \bibinfo{year}{2016}\natexlab{}.
\newblock \showarticletitle{Deep residual learning for image recognition}. In
  \bibinfo{booktitle}{\emph{Proceedings of the IEEE conference on computer
  vision and pattern recognition}}. \bibinfo{pages}{770--778}.
\newblock


\bibitem[Li et~al\mbox{.}(2020)]%
        {TiSAS}
\bibfield{author}{\bibinfo{person}{Jiacheng Li}, \bibinfo{person}{Yujie Wang},
  {and} \bibinfo{person}{Julian~J. McAuley}.} \bibinfo{year}{2020}\natexlab{}.
\newblock \showarticletitle{Time Interval Aware Self-Attention for Sequential
  Recommendation}. In \bibinfo{booktitle}{\emph{{WSDM} '20: The Thirteenth
  {ACM} International Conference on Web Search and Data Mining, Houston, TX,
  USA, February 3-7, 2020}}, \bibfield{editor}{\bibinfo{person}{James
  Caverlee}, \bibinfo{person}{Xia~(Ben) Hu}, \bibinfo{person}{Mounia Lalmas},
  {and} \bibinfo{person}{Wei Wang}} (Eds.). \bibinfo{publisher}{{ACM}},
  \bibinfo{pages}{322--330}.
\newblock


\bibitem[Lin et~al\mbox{.}(2022)]%
        {StEN}
\bibfield{author}{\bibinfo{person}{Shaochuan Lin}, \bibinfo{person}{Yicong Yu},
  \bibinfo{person}{Xiyu Ji}, \bibinfo{person}{Taotao Zhou},
  \bibinfo{person}{Hengxu He}, \bibinfo{person}{Zisen Sang},
  \bibinfo{person}{Jia Jia}, \bibinfo{person}{Guodong Cao}, {and}
  \bibinfo{person}{Ning Hu}.} \bibinfo{year}{2022}\natexlab{}.
\newblock \showarticletitle{Spatiotemporal-Enhanced Network for Click-Through
  Rate Prediction in Location-based Services}. In
  \bibinfo{booktitle}{\emph{Proceedings of the Workshop on Deep Learning for
  Search and Recommendation {(DL4SR} 2022) co-located with the 31st {ACM}
  International Conference on Information and Knowledge Management {(CIKM}
  2022), Atlanta, Georgia, USA, October 17-21, 2022}}
  \emph{(\bibinfo{series}{{CEUR} Workshop Proceedings},
  Vol.~\bibinfo{volume}{3317})}, \bibfield{editor}{\bibinfo{person}{Wei Liu}
  {and} \bibinfo{person}{Linsey Pang}} (Eds.).
  \bibinfo{publisher}{CEUR-WS.org}.
\newblock


\bibitem[Luo et~al\mbox{.}(2021)]%
        {STAN}
\bibfield{author}{\bibinfo{person}{Yingtao Luo}, \bibinfo{person}{Qiang Liu},
  {and} \bibinfo{person}{Zhaocheng Liu}.} \bibinfo{year}{2021}\natexlab{}.
\newblock \showarticletitle{{STAN:} Spatio-Temporal Attention Network for Next
  Location Recommendation}. In \bibinfo{booktitle}{\emph{{WWW} '21: The Web
  Conference 2021, Virtual Event / Ljubljana, Slovenia, April 19-23, 2021}},
  \bibfield{editor}{\bibinfo{person}{Jure Leskovec}, \bibinfo{person}{Marko
  Grobelnik}, \bibinfo{person}{Marc Najork}, \bibinfo{person}{Jie Tang}, {and}
  \bibinfo{person}{Leila Zia}} (Eds.). \bibinfo{publisher}{{ACM} / {IW3C2}},
  \bibinfo{pages}{2177--2185}.
\newblock
\urldef\tempurl%
\url{https://doi.org/10.1145/3442381.3449998}
\showDOI{\tempurl}


\bibitem[Qi et~al\mbox{.}(2021)]%
        {TRISAN}
\bibfield{author}{\bibinfo{person}{Yi Qi}, \bibinfo{person}{Ke Hu},
  \bibinfo{person}{Bo Zhang}, \bibinfo{person}{Jia Cheng}, {and}
  \bibinfo{person}{Jun Lei}.} \bibinfo{year}{2021}\natexlab{}.
\newblock \showarticletitle{Trilateral Spatiotemporal Attention Network for
  User Behavior Modeling in Location-based Search}. In
  \bibinfo{booktitle}{\emph{{CIKM} '21: The 30th {ACM} International Conference
  on Information and Knowledge Management, Virtual Event, Queensland,
  Australia, November 1 - 5, 2021}},
  \bibfield{editor}{\bibinfo{person}{Gianluca Demartini},
  \bibinfo{person}{Guido Zuccon}, \bibinfo{person}{J.~Shane Culpepper},
  \bibinfo{person}{Zi~Huang}, {and} \bibinfo{person}{Hanghang Tong}} (Eds.).
  \bibinfo{publisher}{{ACM}}, \bibinfo{pages}{3373--3377}.
\newblock


\bibitem[Zhou et~al\mbox{.}(2019)]%
        {DIEN}
\bibfield{author}{\bibinfo{person}{Guorui Zhou}, \bibinfo{person}{Na Mou},
  \bibinfo{person}{Ying Fan}, \bibinfo{person}{Qi Pi}, \bibinfo{person}{Weijie
  Bian}, \bibinfo{person}{Chang Zhou}, \bibinfo{person}{Xiaoqiang Zhu}, {and}
  \bibinfo{person}{Kun Gai}.} \bibinfo{year}{2019}\natexlab{}.
\newblock \showarticletitle{Deep Interest Evolution Network for Click-Through
  Rate Prediction}. In \bibinfo{booktitle}{\emph{The Thirty-Third {AAAI}
  Conference on Artificial Intelligence, {AAAI} 2019, The Thirty-First
  Innovative Applications of Artificial Intelligence Conference, {IAAI} 2019,
  The Ninth {AAAI} Symposium on Educational Advances in Artificial
  Intelligence, {EAAI} 2019, Honolulu, Hawaii, USA, January 27 - February 1,
  2019}}. \bibinfo{publisher}{{AAAI} Press}, \bibinfo{pages}{5941--5948}.
\newblock
\urldef\tempurl%
\url{https://doi.org/10.1609/aaai.v33i01.33015941}
\showDOI{\tempurl}


\bibitem[Zhou et~al\mbox{.}(2018)]%
        {DIN}
\bibfield{author}{\bibinfo{person}{Guorui Zhou}, \bibinfo{person}{Xiaoqiang
  Zhu}, \bibinfo{person}{Chengru Song}, \bibinfo{person}{Ying Fan},
  \bibinfo{person}{Han Zhu}, \bibinfo{person}{Xiao Ma},
  \bibinfo{person}{Yanghui Yan}, \bibinfo{person}{Junqi Jin},
  \bibinfo{person}{Han Li}, {and} \bibinfo{person}{Kun Gai}.}
  \bibinfo{year}{2018}\natexlab{}.
\newblock \showarticletitle{Deep Interest Network for Click-Through Rate
  Prediction}. In \bibinfo{booktitle}{\emph{Proceedings of the 24th {ACM}
  {SIGKDD} International Conference on Knowledge Discovery {\&} Data Mining,
  {KDD} 2018, London, UK, August 19-23, 2018}},
  \bibfield{editor}{\bibinfo{person}{Yike Guo} {and} \bibinfo{person}{Faisal
  Farooq}} (Eds.). \bibinfo{publisher}{{ACM}}, \bibinfo{pages}{1059--1068}.
\newblock
\urldef\tempurl%
\url{https://doi.org/10.1145/3219819.3219823}
\showDOI{\tempurl}


\end{thebibliography}

% \begin{table*}
%   \caption{Ablation study 2 .}
%   \input{./table/ablation2.tex}
%   \label{table:ablation2}
% \end{table*}

\end{document}